\documentclass[preprint,prb,english,aps,preprintnumbers,amsmath,amssymb]{revtex4}
\usepackage[T1]{fontenc}
\usepackage[latin9]{inputenc}
\usepackage{verbatim}
\usepackage{float}
\usepackage{graphicx}
\usepackage{subfigure}

\makeatletter

\usepackage{dcolumn} 
\usepackage{bm} 
\usepackage{enumerate}

\DeclareGraphicsExtensions{.jpg,.pdf,.eps,.ps}

\makeatother

\usepackage{babel}

\begin{document}
\title{Simulating lattice spin models on GPUs}
\author{Tal Levy, Guy Cohen, and Eran Rabani} 
\affiliation{School of Chemistry, The Sackler Faculty of Exact
Sciences, Tel Aviv University,
Tel Aviv 69978, Israel}

\begin{abstract}
Lattice spin models are useful for studying critical phenomena and
allow the extraction of equilibrium and dynamical properties.
Simulations of such systems are usually based on Monte Carlo (MC)
techniques and the main difficulty often being the large computational
effort needed when approaching critical points. In this work it is
shown how such simulations can be accelerated with the use of NVIDIA
Graphics Processing Units (GPUs) using the CUDA programming
architecture. We have developed two different algorithms for lattice
spin models, the first useful for equilibrium properties near a second
order phase transition point and the second for dynamical slowing down
near a glass transition. The algorithms are based on parallel MC
techniques and speedups from 70 to 150 fold over conventional
single-threaded computer codes are obtained using consumer-grade
hardware.
\end{abstract}
\maketitle

\section{Introduction} \label{sec:intro}
In most common cases, computer programs are written serially: to solve
a problem, an algorithm is constructed and implemented as a serial
stream of coded instructions. These instructions are executed on a
Central Processing Unit (CPU) on one computer. Momentarily
disregarding specific cases for which modern CPU hardware is
optimized, only one instruction may be executed at a time; on its
termination, the next instruction begins its execution. On the other
hand, parallel computing involves the simultaneous use of multiple
computational resources.  In recent years, massively parallel
computing has become a valuable tool, mainly due to the rapid
development of the Graphics Processing Unit (GPU), a highly parallel,
multi-threaded, many core processor. However, due to the specialized
nature of the hardware, the technical difficulties involved in GPU
programming for scientific use stymied progress in the direction of
using GPUs as general-purpose parallel computing devices. This
situation has begun to change more recently, as NVIDIA introduced
$\text{CUDA}^{\text{TM}}$, a general purpose parallel computing
architecture with a new parallel programming model and instruction set
architecture (similar technology is also available from ATI but was
not studied by us). CUDA comes with a software environment that allows
developers to use C (C++, CUDA FORTRAN, OpenCL, and DirectCompute are
now also supported~\cite{ProgrammingGuide2010}) for program
development, exposing an application programming interface and
language extensions that allow access to the GPU without specialized
assembler or graphics-oriented interfaces. Nowadays, there are quite a
few scientific applications that are running on GPUs; this includes
quantum chemistry
applications~\cite{Ufimtsev+Martinez2008,Ufimtsev2009}, quantum Monte
Carlo simulations~\cite{Anderson2007,Meredith2009}, molecular
dynamics\cite{Meel2007,Stone2007,Anderson2008,Davis2009,Friedrichs2009,Genovese2009,Dematte2010,Eastman2010},
hydrodynamics~\cite{Bernaschi2010}, classical Monte Carlo
simulations~\cite{Lee2009,Tobias+Preis2009,Tobias+Preis2010}, and
stochastic
processes~\cite{Juba2008,Januszewski+Kostur2010,Balijepalli2010}.

In this work we revisit the problem of parallel Monte Carlo
simulations of lattice spin models on GPUs. Two generic model systems
were considered:
\begin{enumerate}
\item The two-dimensional (2D) Ising model~\cite{Onsager1944} serves
  as a prototype model for describing critical phenomena and
  equilibrium phase transitions. Numerical analysis of critical
  phenomena is based on computer simulations combined with finite size
  scaling techniques~\cite{Newman1999}. Near the critical point, these
  simulations require large-scale computations and long averaging,
  while current GPU algorithms are limited to small lattice
  sizes~\cite{Tobias+Preis2009} or to spin $\frac{1}{2}$
  systems~\cite{Tobias+Preis2010}. Thus, in three-dimensions (3D), the
  solution of lattice spin models still remains a challenge in
  statistical physics.
\item The North-East model~\cite{Reiter1992} serves as a prototype for
	describing the slowing down of glassy systems. The
	computational challenge here is mainly to describe
	correlations and fluctuations on very long timescales, as the
	system approaches the glass temperature
	(concentration)~\cite{Reiter1992}.  As far as we know,
	simulations of facilitated models on GPUs have not been
	explored so far, but their applications on CPU has been
	discussed
	extensively~\cite{Garrahan2003,Jack2006,Garrahan2007,Chandler2010a}.
\end{enumerate}
We develop two new algorithms useful to simulate these lattice spin
models. For certain conditions, we obtain speedups of two orders of
magnitude in comparison with serial CPU simulations. We would like to
note that in many GPU applications, speedups are reported in Gflops,
while in the present work, we report speedups of the actual running
time of the full simulation. The latter provides a realistic estimate
of the performance of the algorithms.\\ The paper is organized as
follows: Section~\ref{sec:gpuarch} comprises a brief introduction to
GPUs and summarizes their main features.  Section~\ref{spin models}
contains a short overview of lattice spin models and the algorithms
developed. Section~\ref{PRNG} includes a short discussion of the
random number generator used in this work. Section~\ref{sec:result}
presents the results and the comparisons between CPU and GPU
performance. Finally, Section~\ref{sec:conclusions} concludes.


\section{GPU architecture} \label{sec:gpuarch}
In order to understand GPU programming we provide a sketch of NVIDIA's
GPU device architecture. This is important for the development of the
algorithms reported below, and in particular for understanding the
logic and limitations behind our approach. By now three generations of
GPUs have been released by NVIDIA (G$80$, GT$200$ and the latest
architecture codename ``Fermi''). In table~\ref{tab:3architectures} we
highlight the main features and differences between generations.

\subsection{Hardware specifications} \label{subsec:hardware}
\begin{figure}[h]
\begin{center}
\subfigure[]{\label{fig:Different
philosophies(a)}\includegraphics[width=7cm]{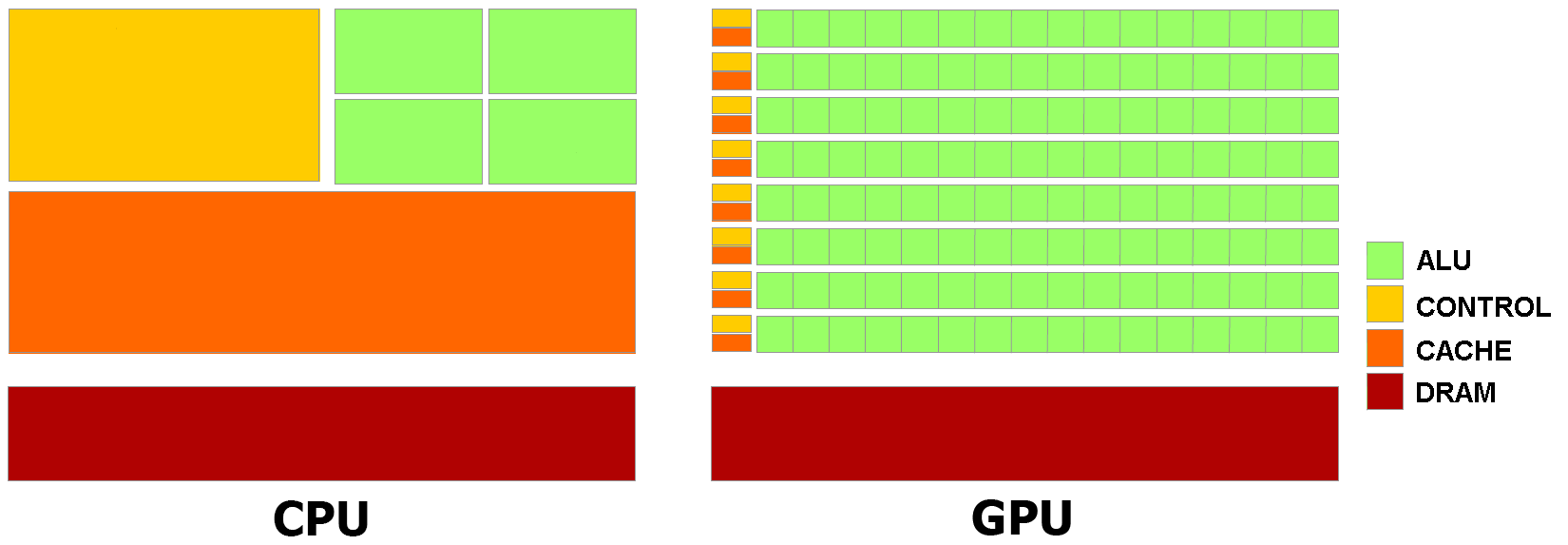}}
\subfigure[]{\label{fig:Different
philosophies(b)}\includegraphics[width=6cm]{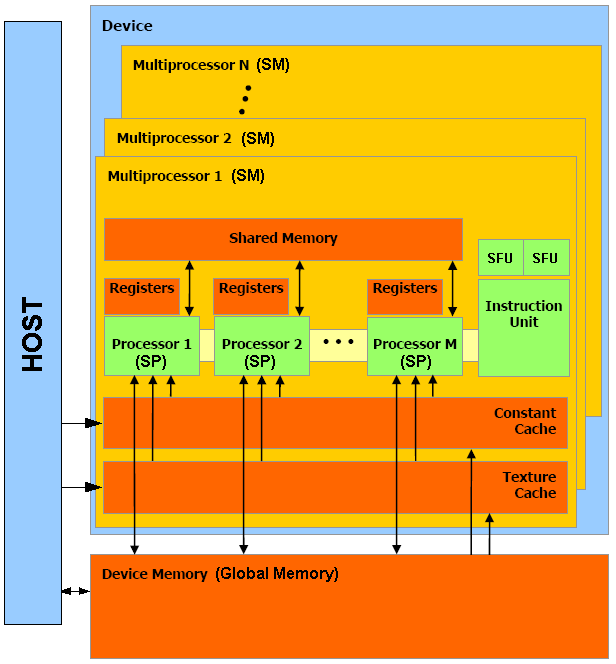}}
\caption{{\small (a) A schematic delineation that illustrates the
ratio of transistors devoted to data processing (ALUs) to those of
data caching and flow control in CPUs and GPUs. This is a
manifestation of the different philosophies between the ``all
purpose'' CPU to the ``dedicated purpose'' GPU. (b) A visual hardware
model that describes the memory architecture and hierarchy of a GPU
device. (Courtesy of NVIDIA).}}
\label{fig:Different philosophies}
\end{center}
\end{figure}

On a GPU device one finds a number of Scalar Multiprocessors (SMs).
Each multiprocessor contains $8/32$ (architecture dependent) Scalar
Processor cores (SPs), a Multi-threaded Instruction Unit (MTIU),
special function units for transcendental numbers and the execution of
transcendental instructions such as sine, cosine, reciprocal, and
square root, $32$-bit registers and the shared memory space. In
general, unlike a CPU, a GPU is specialized for compute-intensive,
highly parallel computation - exactly the task graphics rendering
requires - and is therefore designed such that more transistors are
devoted to data processing rather than data caching and flow
control. This is schematically illustrated in figure
\ref{fig:Different philosophies}, and makes GPUs less general-purpose
but highly effective for data-parallel computation with high
arithmetic intensity. Specifically, they are optimized for
computations where the same instructions are executed on different
data elements (often called Single Instruction Multiple Data, or SIMD)
and where the ratio of arithmetic operations to memory operations is
high. This puts a heavy restriction on the types of computations that
optimally utilize the GPU, but in cases where the architecture is
suitable to the task at hand it speeds up the calculations
significantly.

\subsection{Memory architecture} \label{subsec:memarch}
NVIDIA's GPU memory model is highly hierarchical and is divided into
several layers:
\begin{description}
\item [{Registers:}] The fastest form of memory on the GPU.  Usually
	automatic variables declared in a kernel reside in registers,
	which provide very fast access.
\item [{Local Memory:}] Local memory is a memory abstraction that
	implies ``local'' in the scope of each thread. It is not an
	actual hardware component of the SM. In fact, local memory
	resides in device memory allocated by the compiler and
	delivers the same performance as any other global memory
	region.
\item [{Shared memory:}] Can be as fast as a register when there are
	no bank conflicts or when reading from the same address. It is
	located on the SM and is accessible by any thread of the block
	from which it was created. It has the lifetime of the block
	(which will be defined in the next subsection).
\item [{Global memory:}] Potentially about $400-600$ times slower than
	register or shared memory. This module is accessible by all
	threads and blocks. Data transfers from and to the GPU are
	done from global memory.
\item [{Constant memory:}] Constant memory is read only from kernels
	and is hardware optimized for the case when all threads read
	the same location(i.e. it is cached). If threads read from
	multiple locations, the accesses are serialized.
\item [{Texture memory:}] Graphics processors provide texture memory
	to accelerate frequently performed operations. The texture
	memory space is cached. The texture cache is optimized for
	$2\mbox{D}$ spatial locality, so threads of the same warp that
	read texture addresses that are close together will achieve
	best performance. Constant and texture memory reside in device
	memory.
\end{description}

\subsection{CUDA programing model} \label{subsec:cudaprog}
As has been mentioned, CUDA includes a software environment that
allows developers to develop applications mostly in the C programming
language. C for CUDA extends C by allowing the programmer to define C
functions (\emph{kernels}) that, when called, are executed $N$ times
in parallel by $N$ different CUDA threads on the GPU scalar
processors.  Before invoking a kernel, the programmer needs to define
the number of \emph{threads} ($N$) to be created. Moreover the
programmer needs to decide into how many \emph{blocks} these threads
will be divided. When the kernel is invoked, blocks are distributed
evenly to the different multiprocessors (hereby, having $p$
multiprocessors, requires a minimum of $p$ blocks to achieve $100\%$
utilization of the GPU). The blocks might be $1$, $2$ or $3$
dimensional with up to $512$ threads per block.  Blocks are organized
into a $1$ or $2$ dimensional \emph{grid} containing up to $65,535$
blocks in each dimension. Each of the threads within a block that
execute a kernel is given a unique thread ID. To differentiate between
two threads from two different blocks, each of the blocks is given a
unique ID as well. Threads within a block can cooperate among
themselves by sharing data through shared memory.  Synchronizing
thread execution is possible within a block, but different blocks
execute independently. Threads belonging to different blocks can
execute on different multiprocessors and must exchange data through
the global memory. There is no efficient way to synchronize block
execution, that is, in which order or on which multiprocessor they
will be processed, thus, an efficient algorithm should avoid
communication between blocks as much as possible.\\ Much of the
challenge in developing an efficient GPU algorithm involves
determining an efficient mapping between computational tasks and the
grid/block/thread hierarchy. The multiprocessor maps each thread to
one scalar processor, and each thread executes independently with its
own instruction address and register state. The multiprocessor SIMT
(Single Instruction Multiple Threads) unit creates, manages,
schedules, and executes threads in groups of $32$ called
\emph{warps}. When a multiprocessor is given one or more thread blocks
to execute, it splits them into warps that get scheduled by the SIMT
unit.  The SIMT unit selects a warp that is ready to execute and
issues the next instruction to the active threads of the warp. A warp
executes one common instruction at a time, so full efficiency is
realized when all threads of a warp agree on their execution path (on
the G$80$ and GT$200$ architectures, it takes 4 clock cycles for a
warp to execute, while on the Fermi architecture it takes only one
clock cycle).  If threads of a warp diverge via a data dependent
conditional branch, the warp serially executes each branch path taken,
disabling threads that are not on that path. When all paths terminate,
the threads converge back to the same execution path. Branch
divergence occurs only within a warp: different warps execute
independently regardless of whether they are executing common or
disjointed code paths. More information can be found in
Ref.\cite{ProgrammingGuide2010}.

\begin{table}[ht] 
\begin{tabular}{l c c c}
\hline 
{} & {\small G80} & {\small GT200} & {\small Fermi}\\
\hline
\hline 
{\small Maximum number of multiprocessors} & {\small 16} &
{\small 30} & {\small 16}\\
\hline 
{\small Scalar cores per multiprocessor} & {\small 8} & {\small
8} & {\small 32}\\
\hline 
{\small Double precision capability} & {\small none} & {\small
30 FMA ops/clock$ \ $} & {\small 256 FMA ops/clock}\\
\hline 
{\small Special function units} & {\small 2} & {\small 2} &
{\small 4}\\
\hline 
{\small Warp schedulers per multiprocessor} & {\small 1} &
{\small 1} & {\small 2}\\
\hline 
{\small Shared memory} & {\small 16 KB} & {\small 16 KB} &
{\small 48 KB}\\
\hline 
{\small Concurrent kernels} & {\small 1} & {\small 1} & {\small
Up to 16}\\
\hline
\end{tabular}
\caption{\small Summary of NVIDIA's different architectures}
\label{tab:3architectures}
\end{table}


\section{Lattice spin models} \label{spin models}
We consider a general lattice model of spins that are placed on a
square lattice of dimensions $d=1,2,3...$. The spins may be of any
dimension and may acquire continuous or discrete values. In the
applications reported below, we focus, for simplicity, on the case
where the spins at lattice site $i$ take discrete values of
$s_{i}=\pm1$, but this can be easily extended to any spin dimension
and value. The interactions between the spins is given by the
Hamiltonian:
\begin{equation}
H=\sum_{i\neq j}J_{ij}s_{i}s_{j}+\sum_{i}B_{i}s_{i}.
\end{equation}
In the above equation, the first sum is usually carried over nearest
neighbors only (which we will denote $\langle ij\rangle$).  $J_{ij}$
is the interaction parameter and may be constant, discrete or
continuous and $B_{i}$ is an external field. The above Hamiltonian can
be used to study equilibrium properties as in the Ising model and spin
glass models (Edwards-Anderson model~\cite{Edwards1975},
Sherrington-Kirkpatrick model~\cite{Sherrington1975}, random
orthogonal model~\cite{Marinari1994} etc.), or the dynamic behavior as
in facilitated spin models (Fredrickson and
Andersen~\cite{Fredrickson+Andersen1984}, Jackle-Eisinger north-east
model~\cite{Jackle1991,Reiter1992}, etc.).  Simulations of such
systems are based on Monte Carlo techniques~\cite{Landau2000}. In most
CPU implementations, the algorithms for equilibrium or dynamic
simulations do not differ significantly. However, as will become clear
below, they become very different when implemented on a GPU.  Though
the implementations described below are for two simple cases, the
extension of our approach to the spin glass models mentioned above or
to other facilitated spin models is straightforward.

\subsection{Monte Carlo simulation of the 2D Ising model} \label{subsec:MC2Dising}
\begin{figure}[H]
\begin{center}
\includegraphics[viewport=0bp 380bp 460bp
842bp,clip,height=5cm]{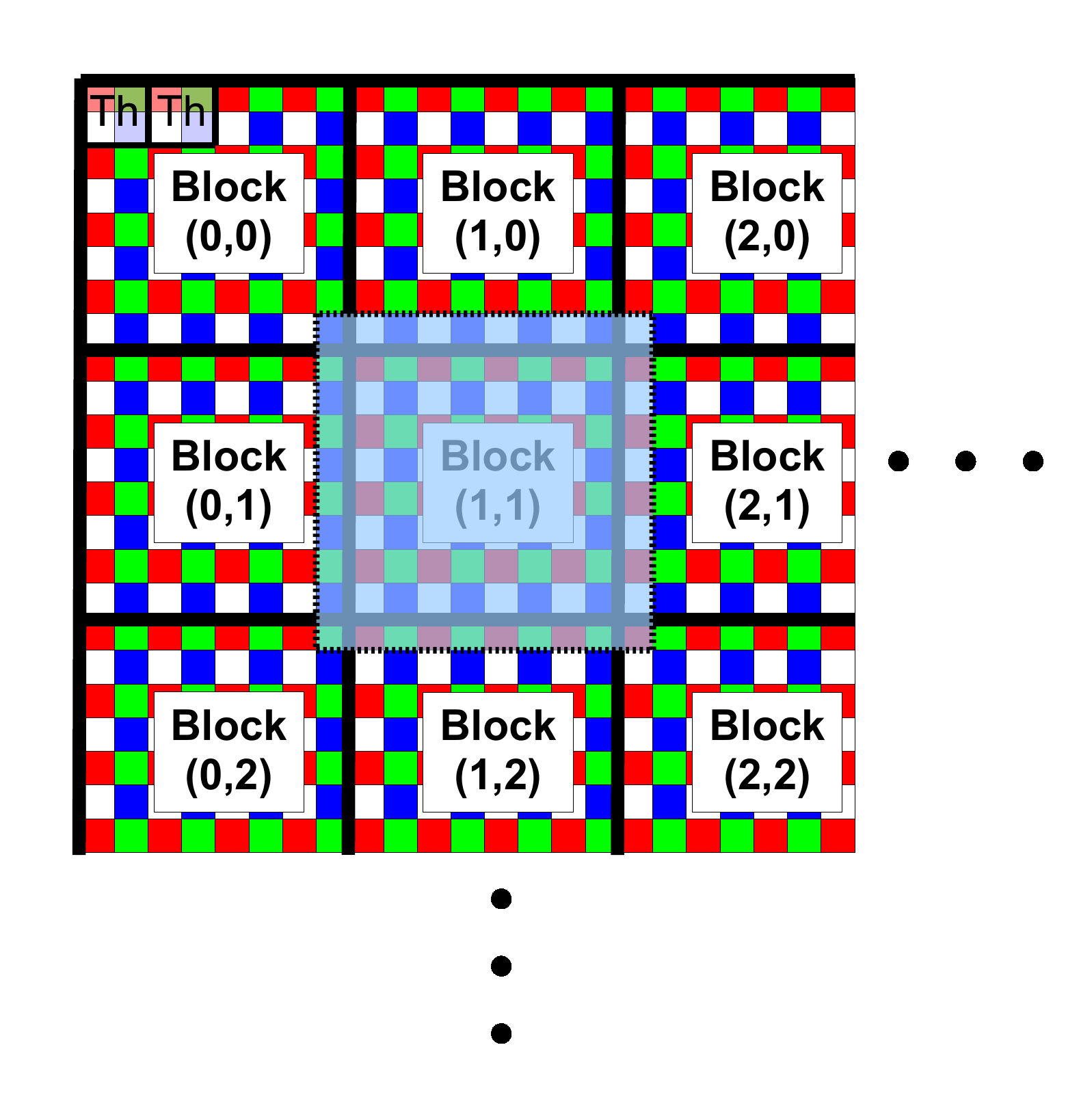}
\caption{{\small Illustration of the algorithm for an Ising model
implemented on a GPU device. Every block handles a sub lattice. Every
thread is assigned $4$ spin sites according to its ID and block. In
order to improve memory access each block copies its sub-lattice on to
shared memory. Since a spin needs its 4 nearest neighbors to update,
extra boundary sites are copied as well (blueish square).}}
\label{fig:2D Ising}
\end{center}
\end{figure}

\begin{figure}[H]
\begin{center}
\subfigure[]{\label{fig:sequence(a)}\includegraphics[scale=0.5]{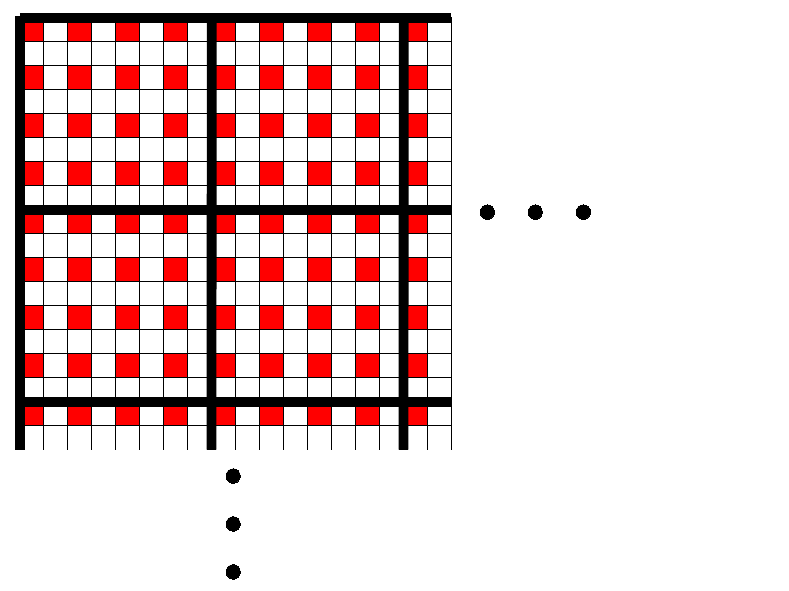}}
\subfigure[]{\label{fig:sequence(b)}\includegraphics[scale=0.5]{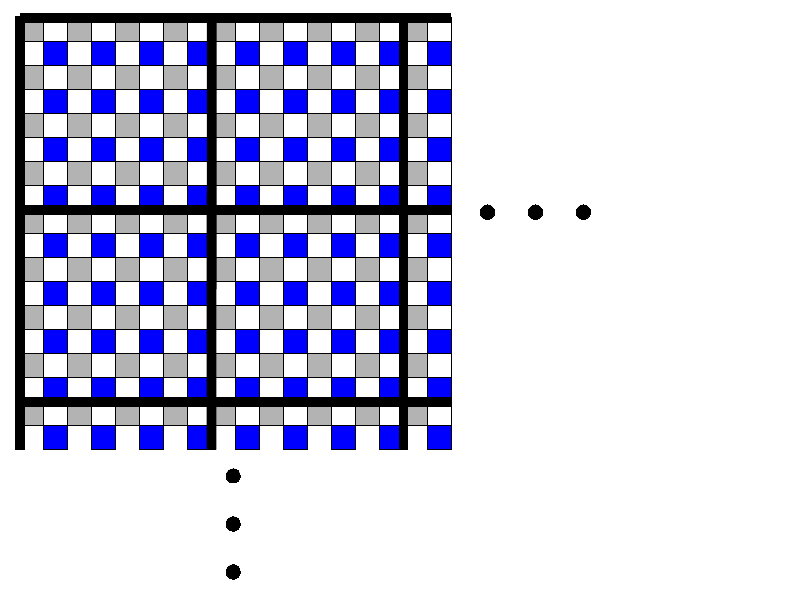}}
\subfigure[]{\label{fig:sequence(c)}\includegraphics[scale=0.5]{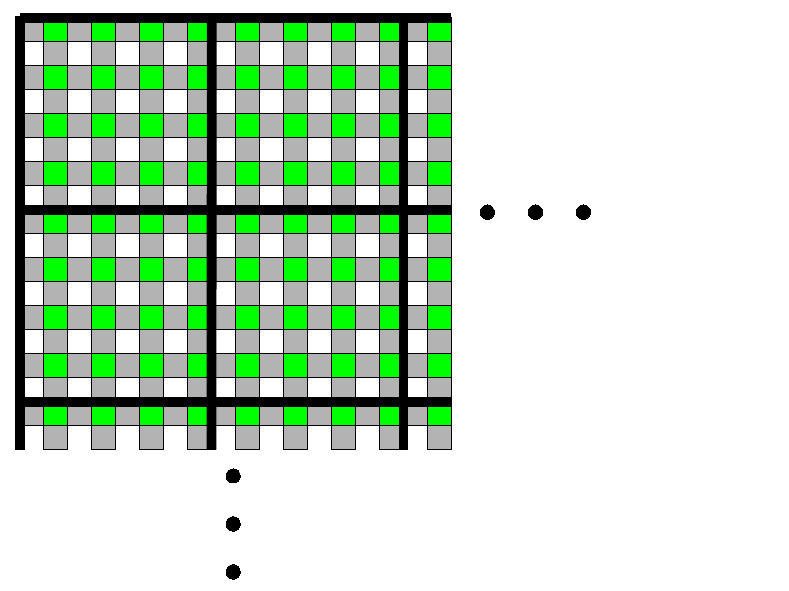}}
\caption{{\small As native block synchronization is not available on a
GPU, one must end the kernel in order to achieve it. This is only
necessary after the red and blue spin where updated (as can be clearly
seen, they do not interact) and before moving on to the green and
white spins, which in turn do interact with the aforementioned updated
red and blue spins.}}
\label{fig:Sequence}
\end{center}
\end{figure}

The simplest $2\mbox{D}$ Ising model~\cite{Newman1999} describes$N$
magnetic dipoles (or spins) placed on a $2\mbox{D}$ square lattice
with one spin per cell. We limit the discussion to the spin
$\frac{1}{2}$ case, where each spin has only two possible
orientations, ``up''and ``down''. Each spin interacts with its nearest
neighbors only, with a fixed interaction strength. In the absence of
an external magnetic field, the Hamiltonian is given by
\begin{equation}
H=-J\sum_{\langle ij\rangle}s_{i}s_{j},
\end{equation}
where $\langle ij\rangle$ represents a sum over nearest neighbors and
$J$ determines the energy scale. Monte Carlo simulation techniques
based on the Metropolis algorithm~\cite{Metropolis+al:1953} are
perhaps the most popular route to obtain the thermodynamic properties
of this model. In a CPU implementation of the Metropolis algorithm, a
spin is selected at random and an attempt to flip the spin is accepted
with the Metropolis probability $P_{acc}=\min[1,\exp(-\Delta H/T)]$,
where $T$ is the temperature in units of energy. In the GPU algorithm
developed here, we will take advantage of the fact that spins interact
only with their nearest neighbors, and thus the problem can be divided
into non-interacting domains. The generalization to the case of finite
interacting regions is straightforward. The algorithm is as follows:
\begin{enumerate}
\item Randomly initialize the lattice (this is done on the CPU).
\item Copy lattice to the GPU.
\item Divide the lattice into $Q$ sub-lattices, each with $P$ spins.
\item A grid of $Q$ thread-blocks is formed.  Every thread-block
	contains $\frac{P}{4}$ threads.
\item Every block copies a sub-lattice including its boundaries from
	the global memory to the shared memory. To avoid bank
	conflicts and save precious shared memory space, the
	\texttt{short} data type was used to form the lattice
	(figure~\ref{fig:2D Ising}).
\item Within the block, every thread is in charge of $4$ spin sites (a
	sub block of $2\times2$). At first all red spins are updated
	(figure~\ref{fig:sequence(a)}), i.e. all threads are
	active. Once a thread finished updating the red spin it
	continues to update its blue spin
	(figure~\ref{fig:sequence(b)}).  Since no native block
	synchronization exists, before updating the remaining spins,
	we make sure that all blocks finished the first two
	steps. This is done by recopying the data from the shared
	memory (boundaries excluded) on to the global memory and
	ending the kernel.
\item Relaunch the kernel with the configuration generated in the
	previous step. Redo steps $5$ and $6$, only this time the
	green and white spins are being updated
	(figure~\ref{fig:sequence(c)}).
\item Recopy data to global memory.
\end{enumerate}
This completes one Monte Carlo step (or one lattice sweep). In our
implementation we chose sub-lattices of $32\times32$ in size and
blocks of $256$ threads, as this choice turned out to be the most
efficient. Using more threads does in fact reduce the time it takes
the block to copy a sub-lattice to the shared memory, but then the
ratio of arithmetic operations to memory operations is low and
performance is poor. To obtain thermodynamic average properties, we
use the fact the CPU and GPU can work in parallel and the lattice is
copied from the device to the host from time to time, so averages can
be calculated on the CPU while the GPU continues to sweep the
lattice. We note in passing that similar algorithms have been proposed
by Tobias {\em et al.}~\cite{Tobias+Preis2009} and Block{\em et
al.}.\cite{Tobias+Preis2010} The former approach is restricted to
lattices with up to $1024\times1024$ spins in 2D (assuming spins are
stored as \texttt{integer} data type), whereas the algorithm presented
in this work is applicable to larger systems, is designed for
coalesced global memory access and avoids bank conflicts, all of which
makes better use of the GPU architecture. The limit of system size in
our approach is related to the size of the global memory on the GPU,
which is typically on the order of $1-4 \mbox{GB}$. This amounts to
maximum system sizes of $20,000\times20,000$ - $40,000\times40,000$
spins. The algorithm of Block {\em et al.}~\cite{Tobias+Preis2010} is
applicable to much larger systems and deals with the issue of
multi-GPU programming, but is currently restricted to spin
$\frac{1}{2}$ systems while the algorithm presented in this work is suitable for the more generalized Potts model\cite{Wu1982Potts} and gives approximately the same speedups in comparison to an equivalent CPU code. 

To acquire the correct thermodynamic equilibrium state, the
chosen set of Monte Carlo moves must satisfy either detailed balance or
the weaker balance condition. On the CPU detailed balance is
rigorously satisfied when randomly selecting a spin within the Metropolis
algorithm. For the GPU, however, the approach we developed breaks detailed
balance and only balance is satisfied~\cite{Manousiouthakis1999,Ren2006}.
This is a sufficient condition to ensure that the algorithm
describes the correct Boltzmann distribution.

\subsection{The North-East model} \label{ne model}
\begin{figure}[ht]
\begin{center}
\includegraphics[viewport=0bp 400bp 540bp
842bp,clip,scale=0.35]{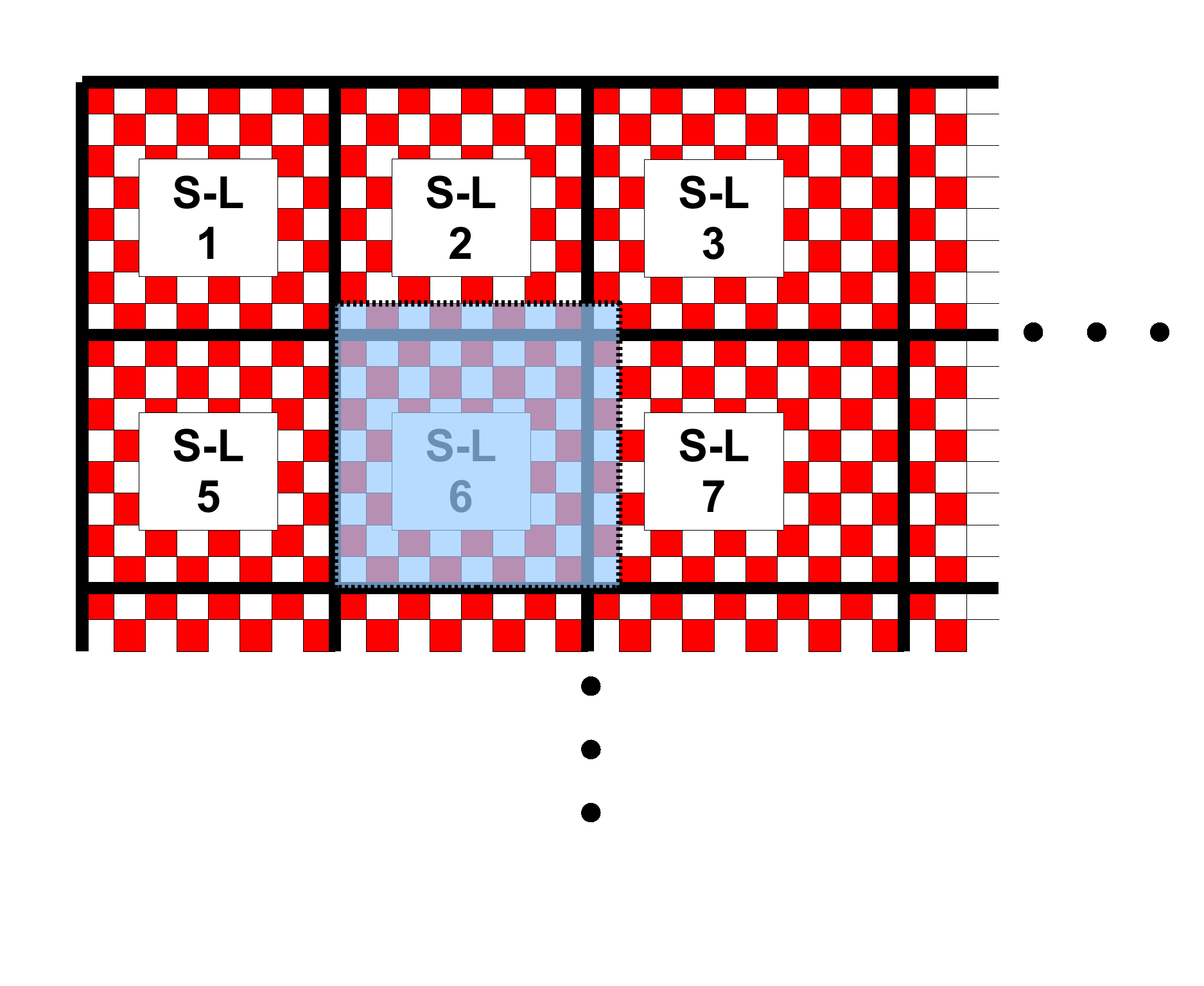}\caption{{\small Although the figure above
looks similar to the Ising model implementation, in this case
Sub-lattices and spins are chosen randomly. This preserves the
dynamical behavior in comparison to CPU algorithms. Note that when
copying the data to the shared memory, only the upper and right
boundaries are necessary.}}
\label{fig:NE}
\end{center}
\end{figure}

\begin{figure}[ht]
\begin{center}
\includegraphics[viewport=0bp 0bp 792bp
605bp,clip,scale=0.3]{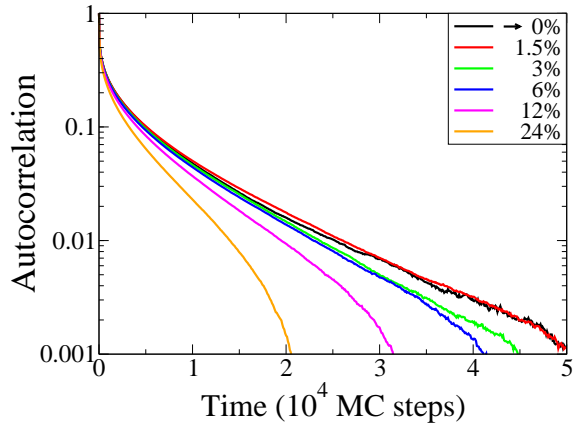}
\caption{{\small Convergence tests of the parameter $\lambda$ for the
North-East model at $c=0.42$.  The solid black line represents the
reference value, which was obtained by running the simulation on a
single thread - single block. This result is identical to the serial
CPU result. The best performence is achieved for sub-lattices of
$32\times32$ spins and for $\lambda=16$ threads, i.e., attempts to
flip only $\sim1.5\% \approx \frac{16}{32\times32}$ of the spins
inside a sub-lattice. One can see that for higher values of $\lambda$,
the correlation diverges from the correct result.}}
\label{fig:Con of parameter}
\end{center}
\end{figure}

The North-East model is based on the Ising model Hamiltonian with
special constraints that are used to model facilitated
dynamics~\cite{Fredrickson+Andersen1984,Nakanish+Takano1986,Reiter1992}.
The Hamiltonian is given by:
\begin{equation}
H=J\sum_{\left\langle i,j\right\rangle
}s_{i}s_{j}+B\sum_{i}s_{i},
\end{equation}
where $s_{i}$, $J$, $B$ and $\langle ij\rangle$ are described
above. What makes this model different from the previous one is a
constraint imposed on the transition probability to flip a spin. This
probably is zero unless the spin's upper (north) and right (east)
neighbors point ``up''. In the latter case, one accepts a flip with
the same Metropolis probability given by $P_{acc}=\min[1,\exp(-\Delta
H/T)]$.  Thus, the thermodynamics of the model are the same as the
Ising model, but the Monte Carlo dynamics generated by the above rule
is quite different. In most applications reported in the literature
one takes $J=0$. For $J\ne0$ the dynamics generated by this model are
richer and show an interesting re-entrant
transition~\cite{Geissler2005}. When $J=0$ the thermodynamics are
trivial and the coupling between neighboring spins depends only on the
aforementioned dynamical constraint. The lattice is initialized so
spins point ``up'' with the probability $c$, which is also the
equilibrium density of spins pointing {}``up'', and can be expressed
as $c=\frac{1}{z}e^{-1/T}$, where the partition function is
$z=e^{1/T}+e^{-1/T}$ and $T$ is the unit-less temperature. Due to the
dynamical constraints, if $c$ is too low, one finds domains of spins
that are stuck.  For a high values of $c$ on the other hand, all spins
are flippable. As a consequence, there is a critical concentration
$c^{*}$ below which the system is not ergodic. This transition from
ergodic to non-ergodic behavior is modeled by the spin-spin
autocorrelation function
\begin{equation}
\Phi(t)=\sum_{i}\frac{\left\langle s_{i}(t)\cdot
s_{i}(0)\right\rangle -\left\langle s_{i}\right\rangle
^{2}}{1-\left\langle s_{i}\right\rangle ^{2}},
\end{equation}
where $\langle s_{i}\rangle=2c-1$ is the average spin polarization and
$s_{i}(t)$ is the spin polarization at Monte Carlo step $t$ for site
$i$. We expect the function to decay to zero for an initial
concentration $c>c^{*}$ and to decay to a finite value $f$ (which is
the fraction of spins that are stuck) for $c<c^{*}.$ The CPU
implementation of the North-East model is identical to that described
for the Ising model, with the additional constraint for the flipping
probability.\\ The GPU implementation for this model, similar as it
may seem to the Ising model, is a bit cannier. We found out that
applying the checkerboard algorithm (described in
subsection~\ref{subsec:MC2Dising}) does not yield the same relaxation
dynamics as the serial CPU implementation. This is understandable,
since this model imitates a diffusion process: for a spin to be able
to change its configuration it is necessary that its north and east
neighbors point up. If they do not, they in turn will also need their
neighbors to point up to be able and change their configuration.
Equilibration takes place by an up spin diffusion from
{}``north-east'' to {}``south-west''. By sequentially (instead of
randomly) sweeping the lattice we change the dynamics of this
process. Such being the case, the corrected algorithm we developed is:
\begin{enumerate}
\item Initialize the lattice so spins point up with probability $c$
        (on the CPU).
\item Copy lattice to the GPU.
\item Divide the lattice into $Q$ sub-lattices, each with $P$ spins.
\item A grid of $Q$ thread-blocks is formed. Every thread-block
	contains $\frac{P}{4}$ threads.
\item Blocks then randomly pick a sub-lattice (figure~\ref{fig:NE}) in
	such a way that two different blocks cannot pick the same
	sub-lattice.  Every block copies a sub-lattice including its
	boundaries from the global memory to the shared memory.
\item Within each block, $\lambda$ threads concurrently pick $\lambda$
	spins from the sub-lattice randomly and update them. $\lambda$
	is chosen to be a small fraction of $P$.
\item Synchronize the block.
\item Reiterate steps $6-7$ $q$ times, such that $q\cdot\lambda=\frac{P}{2}$.
\item Copy back data from shared memory to global memory (to allow block synchronization).
\item Relaunch the kernel with the same configuration and redo steps $5-8$.
\item Recopy data to global memory.
\end{enumerate}
This completes one lattice sweep. Again we use the fact the CPU and
GPU can work in parallel, and the lattice is copied from the device to
the host from time to time to store the spin's configuration for the
computation of the autocorrelation function.\\ The algorithm presented
here does not preserve detailed balance nor the weaker balance
condition, but since we are interested in its dynamics the given
algorithm is correct. One should note that in order to obtain the
correct dynamics an initialization of the parameter $\lambda$ is
needed. It is obvious that in the limit where $Q=\lambda=1$ the
algorithm is in fact serial and preserves detailed balance (results
for this limit are given by the black line in figure~\ref{fig:Con of
parameter}). We can now use this result as a reference, and increase
the number of threads and blocks that are working in parallel. There
is an upper limit above which the results will diverge from the
desired reference and the dynamics will no longer be correctly
reproduced. Once $\lambda$ has been evaluated, the simulation can be
performed. In figure~\ref{fig:Con of parameter} we provide a
consistent test on the value of $\lambda$ for the North-East model at
$c=0.42$.

It is possible to write an algorithm that maintains the weaker balance
condition and at the same time preserve the dynamical relaxation of
the model. The drawback of such an algorithm, however, is that it is
$\approx 2.5\times$ slower then the one presented above. The only
changes are in steps $6-9$:
\begin{enumerate}
\item [6.]\setcounter{enumi}{6}Within each block, $\lambda$ threads concurrently
	pick $\lambda$ \emph{red} spins (figure~\ref{fig:NE}) from the 	sub-lattice randomly
	and update them. $\lambda$ is chosen to be a small fraction of $P$. 
\item Synchronize the block and pick $\lambda$ \emph{white} spins from
	the sub-lattice randomly and update them.
\item Copy back data from shared memory to global memory (to allow block synchronization).
\item Relaunch the kernel with the same configuration and redo steps $6-8$
	three more times (this will complete one MC step).
\end{enumerate}


\section{Pseudo random numbers generation} \label{PRNG}
Monte Carlo simulations rely heavily on the availability of random or
pseudo-random numbers.  In this work, random numbers were used to
determine whether spin flips are accepted or rejected, in accordance
with the Metropolis algorithm. As is widely known, the use of a poor
quality PRNG (Pseudo Random Numbers Generator) may lead to inaccurate
simulations.\cite{Ferrenberg1992} In the course of this work three
different PRNGs were utilized:
\begin{enumerate}
\item L'Ecuyer with Bays-Durham shuffle and added safeguards.\cite{NR}
	This PRNG has a long period ($2 \times 10^{18}$) and is easy
	to implement on a CPU.  Unfortunately, porting it to GPUs
	causes a dramatic decrease in performance: this is mostly due
	to the fact that the implementation of this algorithm requires
	too many registers.
\item The "minimal" random number generator of Park and
 	Miller.\cite{NR} This PRNG has a period of $2 \times 10^9$ and
 	is portable to GPUs. In practice, however, it proved to work
 	poorly and the GPU simulation results were in poor agreement
 	with reference values obtained with better PRNGs.
\item Linear Congruential Random Number Generator (LCRNG).\cite{NR}
	This PRNG has a period of $10^6-10^9$ and provided good
	results even for very long runs. The LCRNG algorithm is very
	easy to port to the GPU and has the advantage of being very
	fast, requiring only a few operations per call.
\end{enumerate}	
	
In our implementations (section~\ref{spin models}), each thread used a
separate LCRNG, creating its own sequence of pseudo-random numbers
with a unique seed. The sequence (of thread $i$) is created as
follows:
\begin{equation}
x_{j+1}^{Th_i} = \left(a \cdot x_j^{Th_i}+c\right) \mbox{\textbf{mod}} \ m 
\end{equation}
\begin{equation}
x_{j+1}^{Th_i} = \mbox{abs}\left(\frac{x_{j+1}^{Th_i}}{2^{31}} \right)
\end{equation}
An appropriate choice of the coefficients is responsible for the
quality of the LCRNG. We used~\cite{NR} $a=1,664,525$ ,
$c=1,013,904,223$ and $m=2^{32}$. The different seeds were created per
thread according to
\begin{equation}
x_0^{Th_{i+1}} = \left(16807 \cdot x_0^{Th_i} \right) \mbox{\textbf{mod}} \ m
\end{equation}
with $x_0^{Th_0} = 1$. Similar PRNGs were used for other GPU
applications.\cite{Tobias+Preis2009,Tobias+Preis2010}


\section{Results} \label{sec:result}
For comparison, CPU codes were executed on a PC with Intel Core2 Duo
E7400 @ $2.8G\mathrm{Hz}$ processor, Intel RaisinCity motherboard with
Intel G41 chipset and Kingstone $2\mathrm{GB},\,800\mathrm{MHz}$ RAM
(only a single core was used for the calculations).  The operating
system was CENTOS. Codes were compiled with Intel C++ compiler (ICC)
using all optimizations provided for best performance. Speedups
reported here are given in terms of the complete application running
time (from initialization till results are processed), rather than
Gflops or spin updates per second. This choice is important, as it
most closely describes the {}``real'' gain in practical simulations by
the use of a GPU rather than a CPU.

\subsection{Ising model} \label{subsec:ising results}
\begin{figure}[H]
\begin{center}
\subfigure[]{\includegraphics[viewport=0bp 0bp 695bp
612bp,width=5cm]{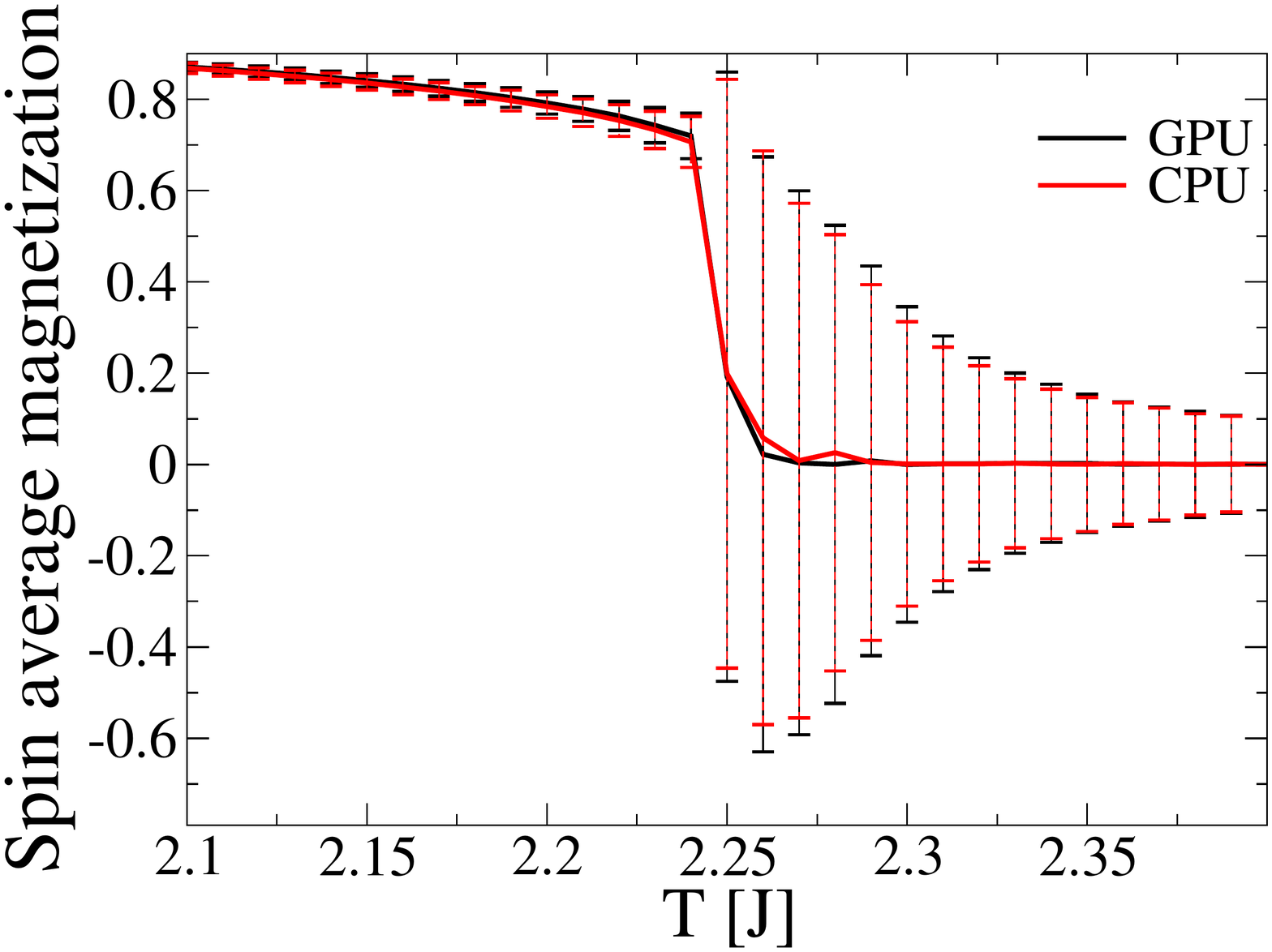}}
\subfigure[]{\includegraphics[viewport=0bp 0bp 695bp
612bp,width=5cm]{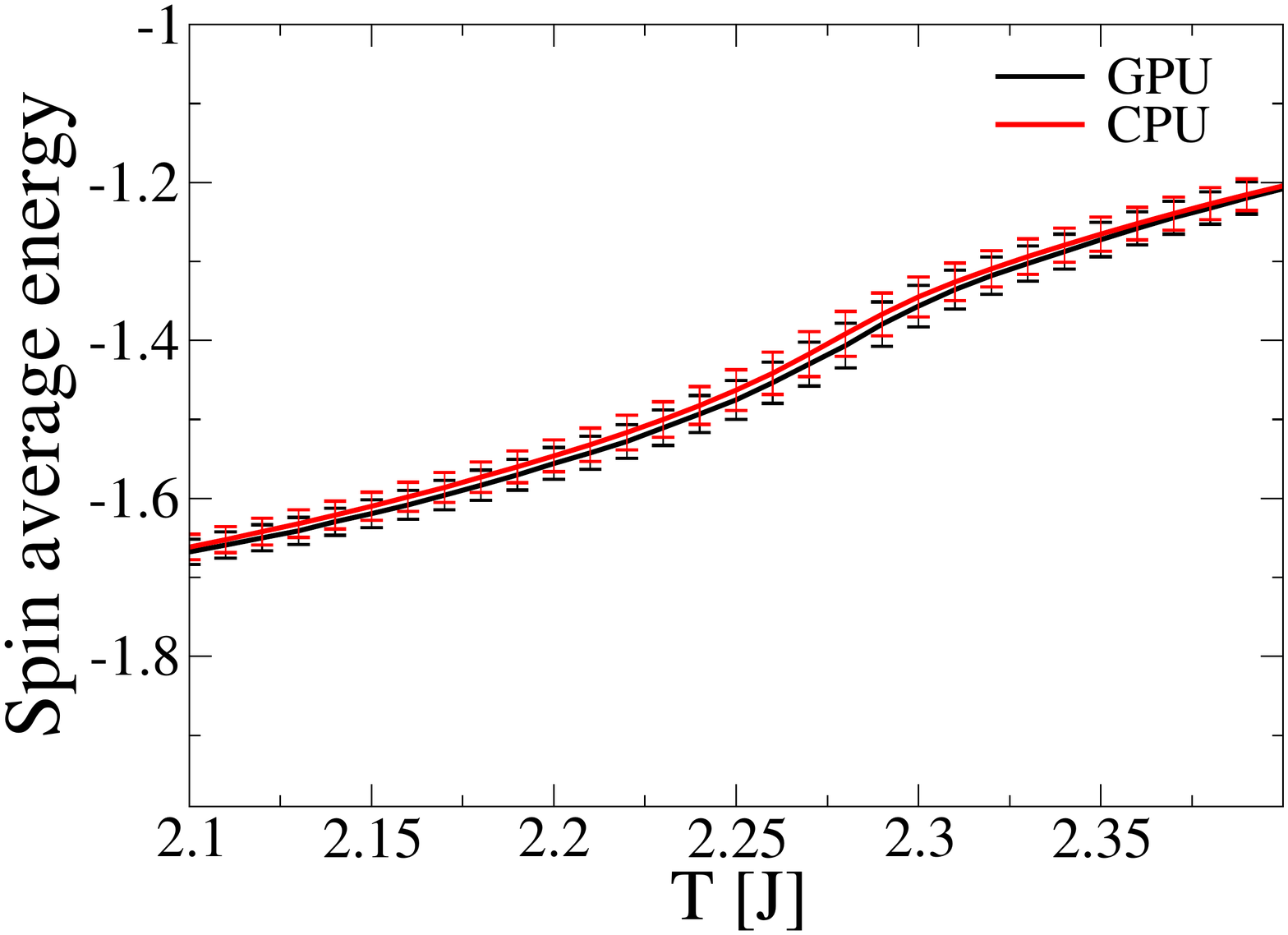}}
\subfigure[]{\includegraphics[viewport=0bp 0bp 690bp
612bp,width=5cm]{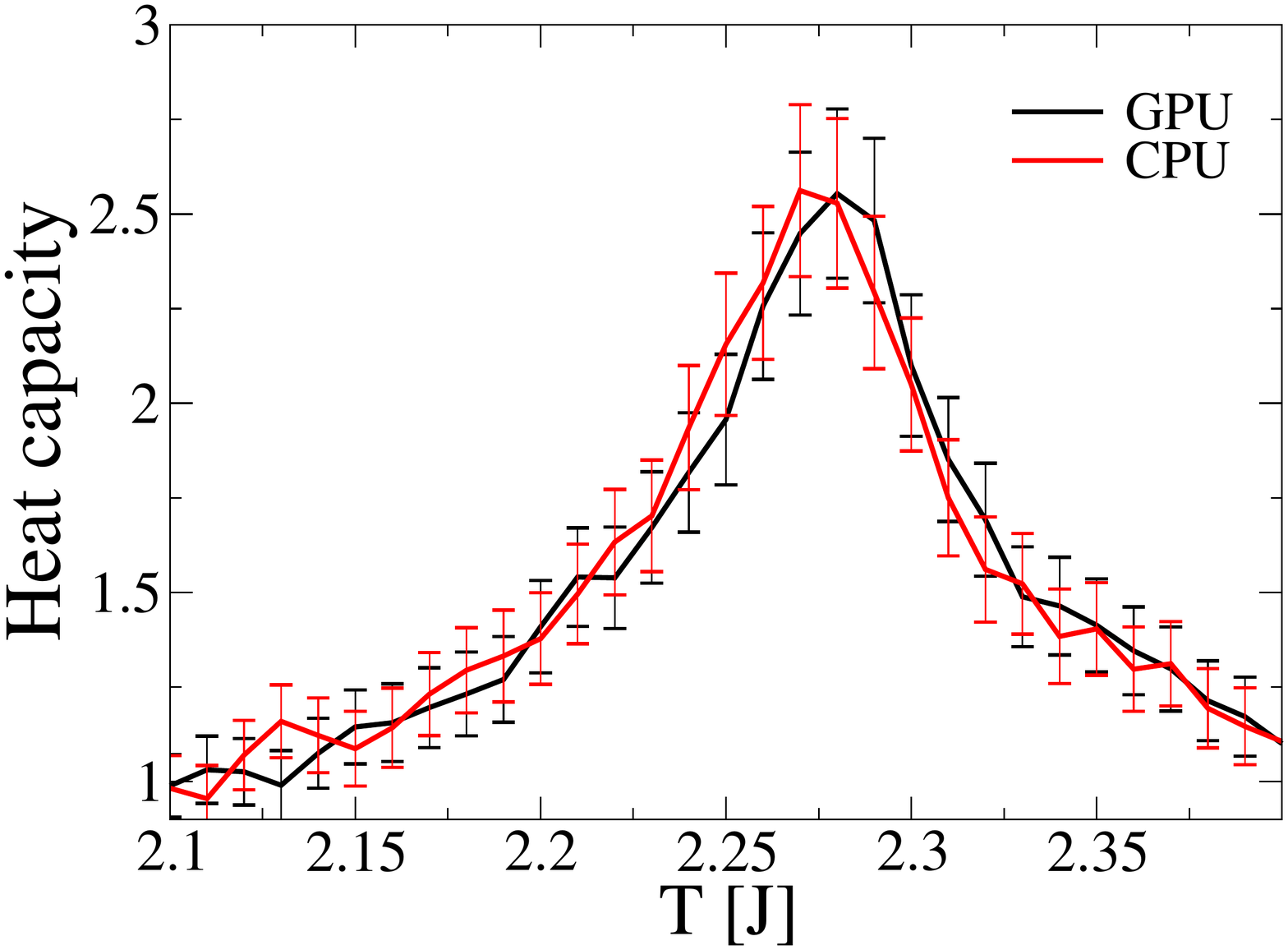}}
\caption{{\small Comparison between the Monte Carlo simulation done on
a CPU and on a GPU for a $128\times128$ spin lattice. The temperature
was stepwise reduced by $0.01$ from $T=2.4$ to $T=2.1$. The critical
temperature for this model is $T_{c}=2/\log(1+\sqrt{2})$. At each
temperature, $2\times10^{7}$ sweeps through the lattice were
performed, during which $10,000$ different measurement were taken
(after reaching equilibrium).  In (a) we show the average spin
magnetization, in (b) the average spin energy and in (c) the heat
capacity.}}
\label{fig:Comparison cpu vs gpu}
\end{center}
\end{figure}

In order to verify the GPU implementation, we compared values of
magnetization, energy and heat capacity as a function of temperature
(temperature was taken in energy units) between GPU and CPU
versions. In figure~\ref{fig:Comparison cpu vs gpu} results from a
$128\times128$ spin lattice are presented. We find the results to
clearly agree. In terms of acceleration we achieved a $15\times$
factor for lattices with $512\times512$ spins on the GT $9600$ GPU
(G$80$ architecture). Implementing the same code on the new GTX $480$
GPU, we achieved a factor of $150\times$. This factor reduces to
$2\times$ for small lattices ($32\times32$).  The reason for this is
that in small problems it becomes harder to hide memory access
latencies, because there are not enough threads to execute between
memory access operations. A theoretical analysis of our implementation
shows that the GPU reaches full occupancy (a useful tool to check this
is the ``CUDA GPU Occupancy Calculator'' which can be freely
downloaded from Ref.\cite{Occupancy}). The $10$ fold factor obtained
by the GTX $480$ in comparison with the GT $9600$ is easily understood
when taking into consideration that the GTX $480$ has$15$ SMs instead
of $8$ (on the GT $9600$), $4$ blocks can be active simultaneously
instead of $3$ and a warp executes in one clock cycle instead of
$4$. Thus the relative speedup is
$\frac{15}{8}\cdot\frac{4}{3}\cdot\frac{4}{1}=10$.  From
figure~\ref{fig:Running times(a)} it is obvious that the GPU reaches
full occupancy only for lattices bigger than $128\times128$. Note also
that near the critical temperature, the fluctuations and noise
increase. Since the present work is not concerned with determining the
critical behavior, we have used the same number of Monte Carlo sweeps
for all temperatures. Estimation of the critical behavior requires
much longer runs, and perhaps also larger systems. In this respect,
the GPU approach developed here provides the means to increase
numerical accuracy by more than one order of magnitude (noise scales
with the square root of the number of MC steps), either by increasing
the system size or by simulating longer Monte Carlo runs.

\subsection{North-East model} \label{subsec:ne results}
The North-East model has a critical concentration $c^{*}\approx0.3$,
below which the dynamics break ergodicity. In figure~\ref{fig:N-E
results} we show the results for spin-spin autocorrelation function
for different concentrations above the critical value. Two lattice
sizes were studied. Similar to the previous case, we find that the GPU
results agree well with the CPU results, indicating that the proposed
algorithm reproduces the correct dynamics. This is not trivial and
depends on the value chosen for $\lambda$. Moreover, as pointed out
above, the algorithm used for the Ising model fails to produce correct
relaxation times. As the system approached the critical concentration
the dynamics become sluggish and the autocorrelation function decays
slowly to zero.
\begin{figure}[H]
\begin{center}
\subfigure[{\footnotesize $c=0.32$,
Time$=10^{8}$}]{\includegraphics[viewport=0bp 0bp 720bp
612bp,clip,width=5cm]{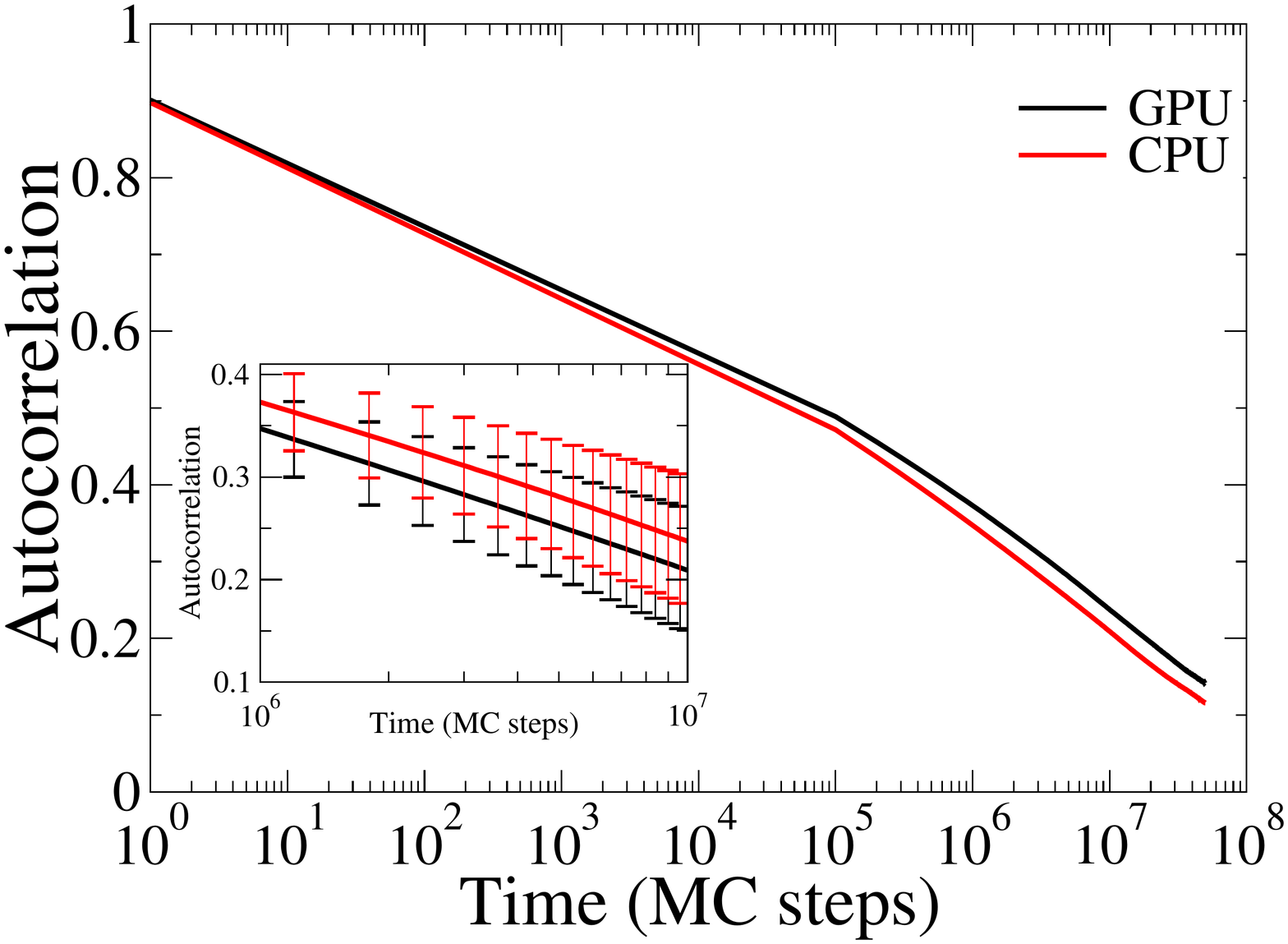}}
\subfigure[{\footnotesize $c=0.37$,
Time$=10^{6}$}]{\includegraphics[viewport=0bp 0bp 720bp
612bp,clip,width=5cm]{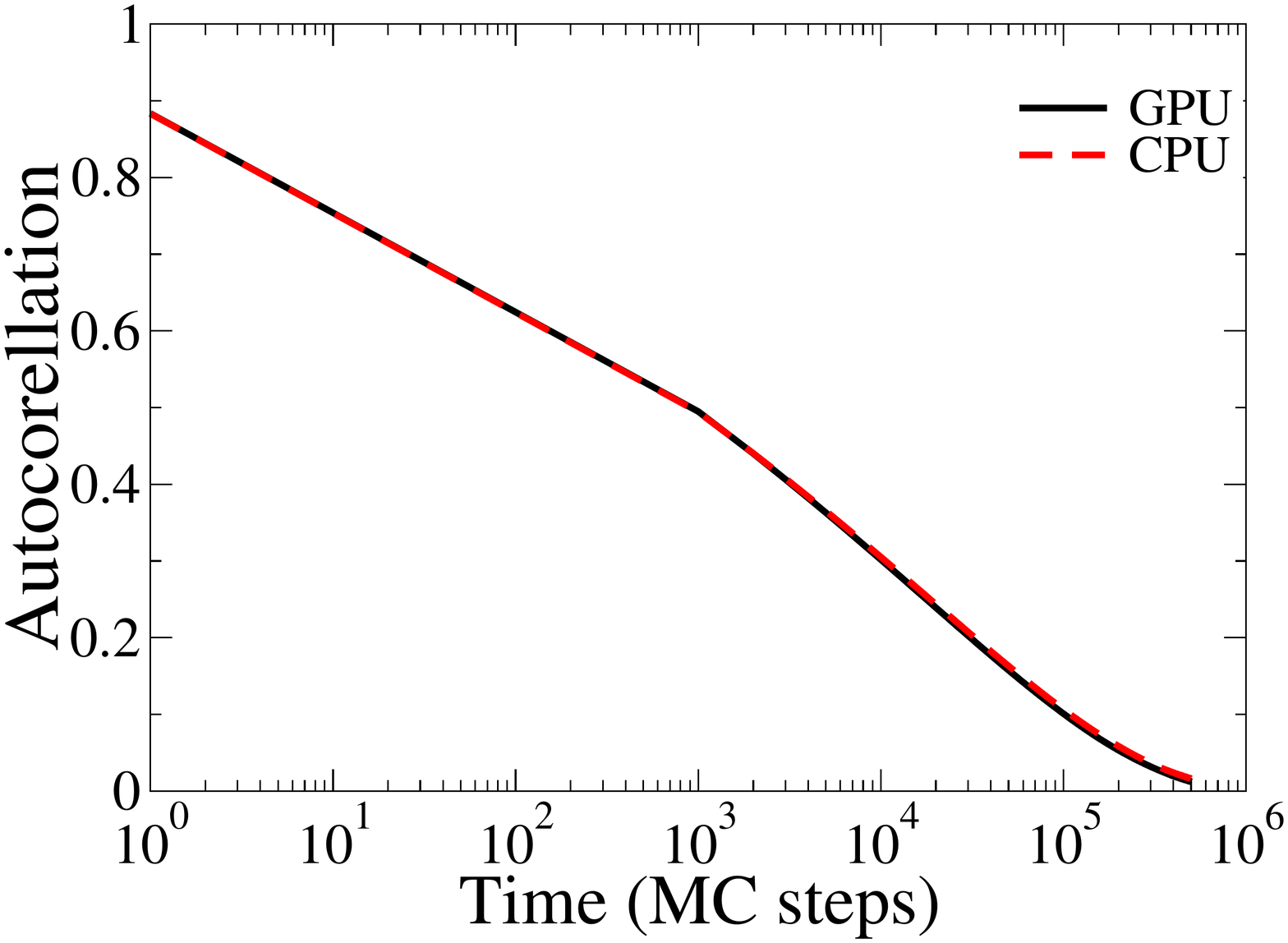}}
\subfigure[{\footnotesize $c=0.42$,
Time$=10^{5}$}]{\includegraphics[viewport=0bp 0bp 720bp
612bp,clip,width=5cm]{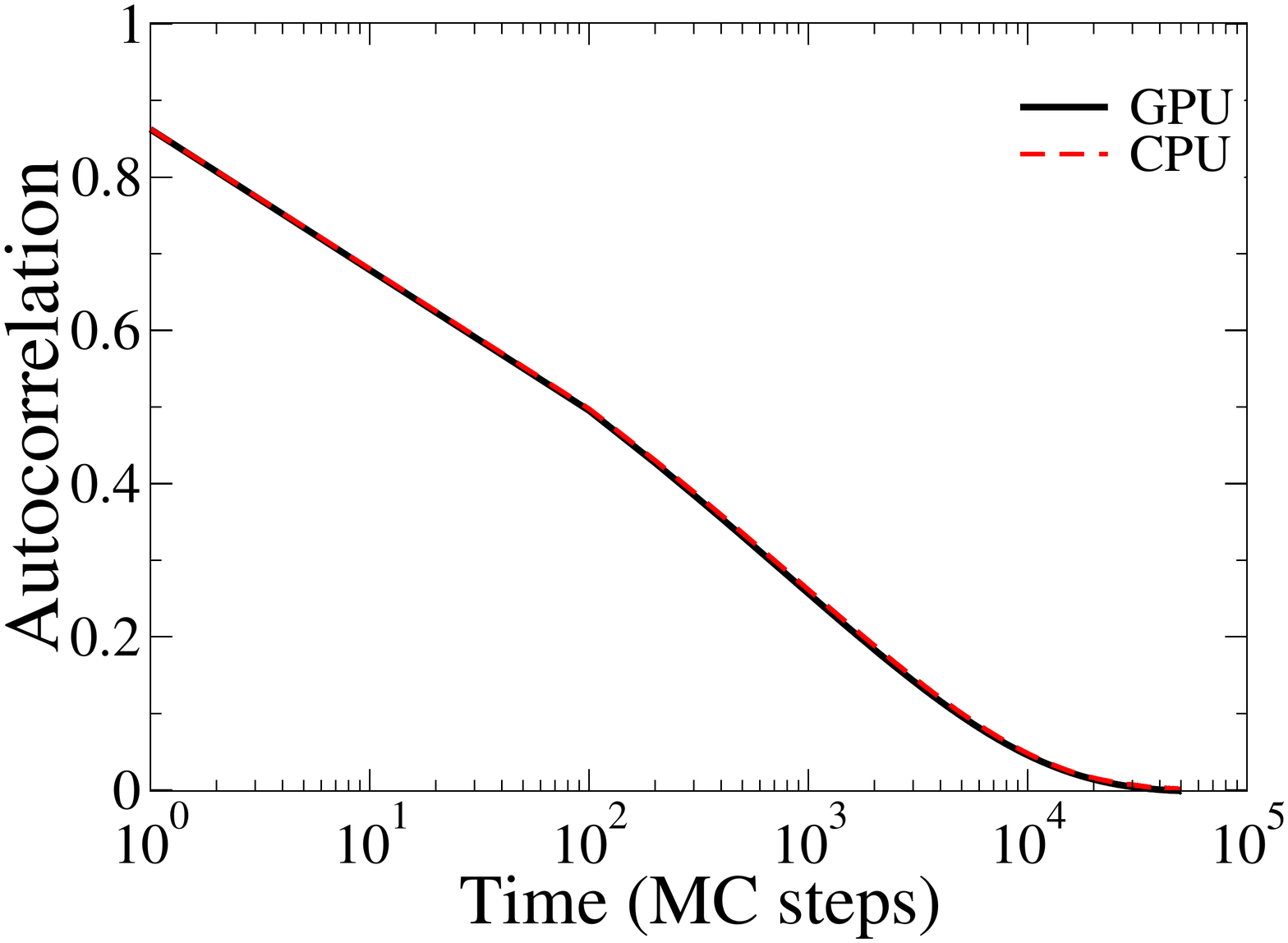}}
\caption{{\small Comparison between spin-spin autocorrelation
functions calculated from simulations done on a CPU and on a GPU for
spin lattices of different sizes and different initial
concentrations. Results were obtained by averaging the autocorrelation
functions of $30$ different initial realization for every
concentration. (a) Results for a $128\times128$ spin lattice. As can
be seen for low concentrations approaching $c^{*}$, relaxation time is
very long (longer than the time we were willing to wait for the CPU
results). Furthermore, at this concentration, more than 30 trajectory
are required to average the autocorrelation function. (b) and (c)
Results obtained from $512\times512$ spin lattices.}}
\label{fig:N-E results}
\end{center}
\end{figure}

In figure~\ref{fig:Running times(b)} we compare the running times
between the CPU and two GPU architectures. We achieved a $2\times$
factor on the GT $9600$ GPU and a $\approx70\times$ speedup running
the same code on the new GTX $480$ GPU. This factor reduces to
$8\times$ for smaller lattices ($128\times128$ and less). The GPU
acceleration is nearly two orders of magnitude, implying that one can
simulate the system closer to the critical density. This is important,
since the behavior of relaxation near the critical density is not
necessarily universal, and thus extrapolations are often tricky.

\begin{figure}[ht]
\begin{center}
\subfigure[]{\label{fig:Running
times(a)}\includegraphics[viewport=0bp 0bp 745bp
612bp,clip,width=6cm]{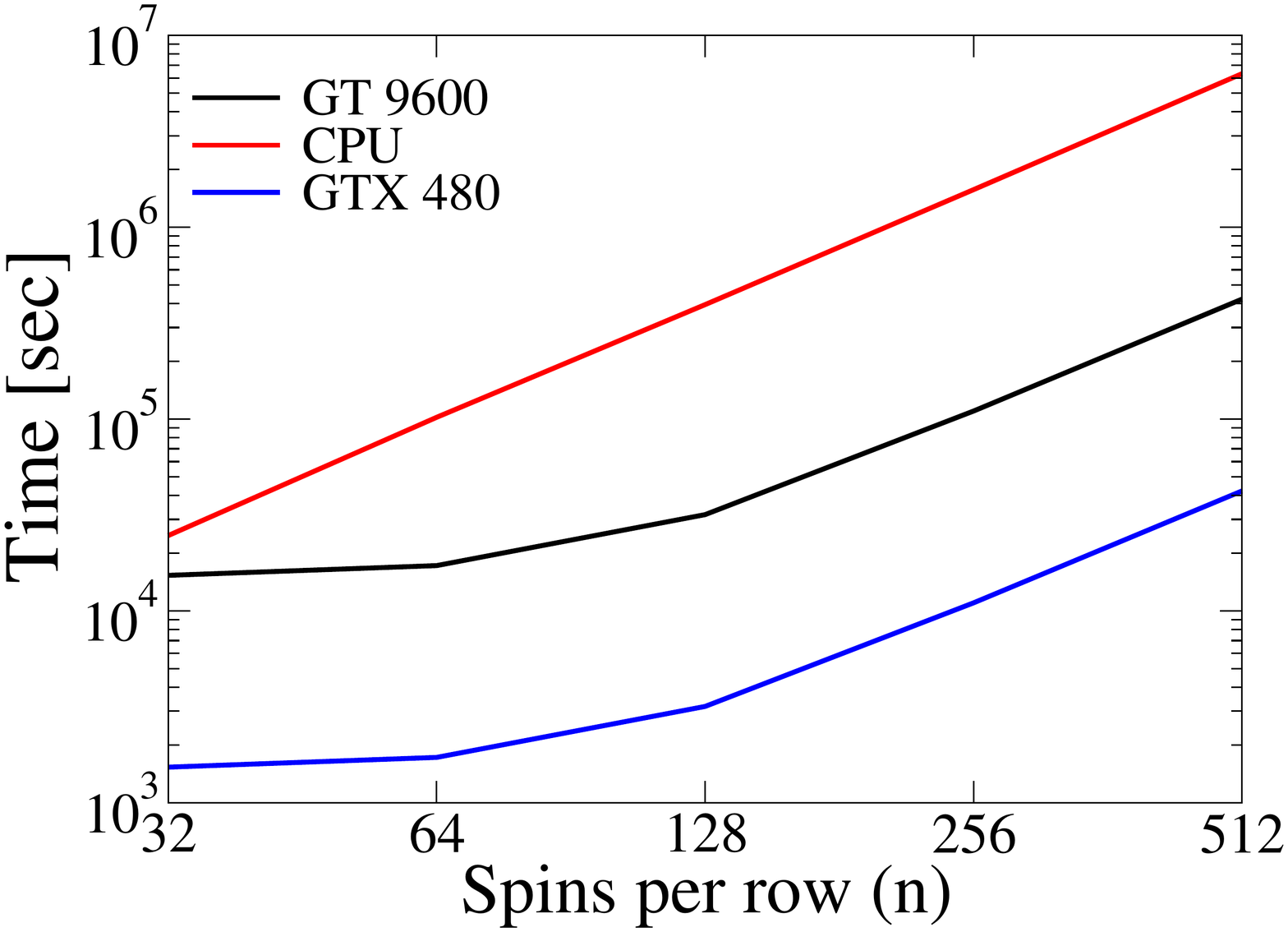}}
\subfigure[]{\label{fig:Running
times(b)}\includegraphics[viewport=0bp 0bp 745bp
612bp,clip,width=6cm]{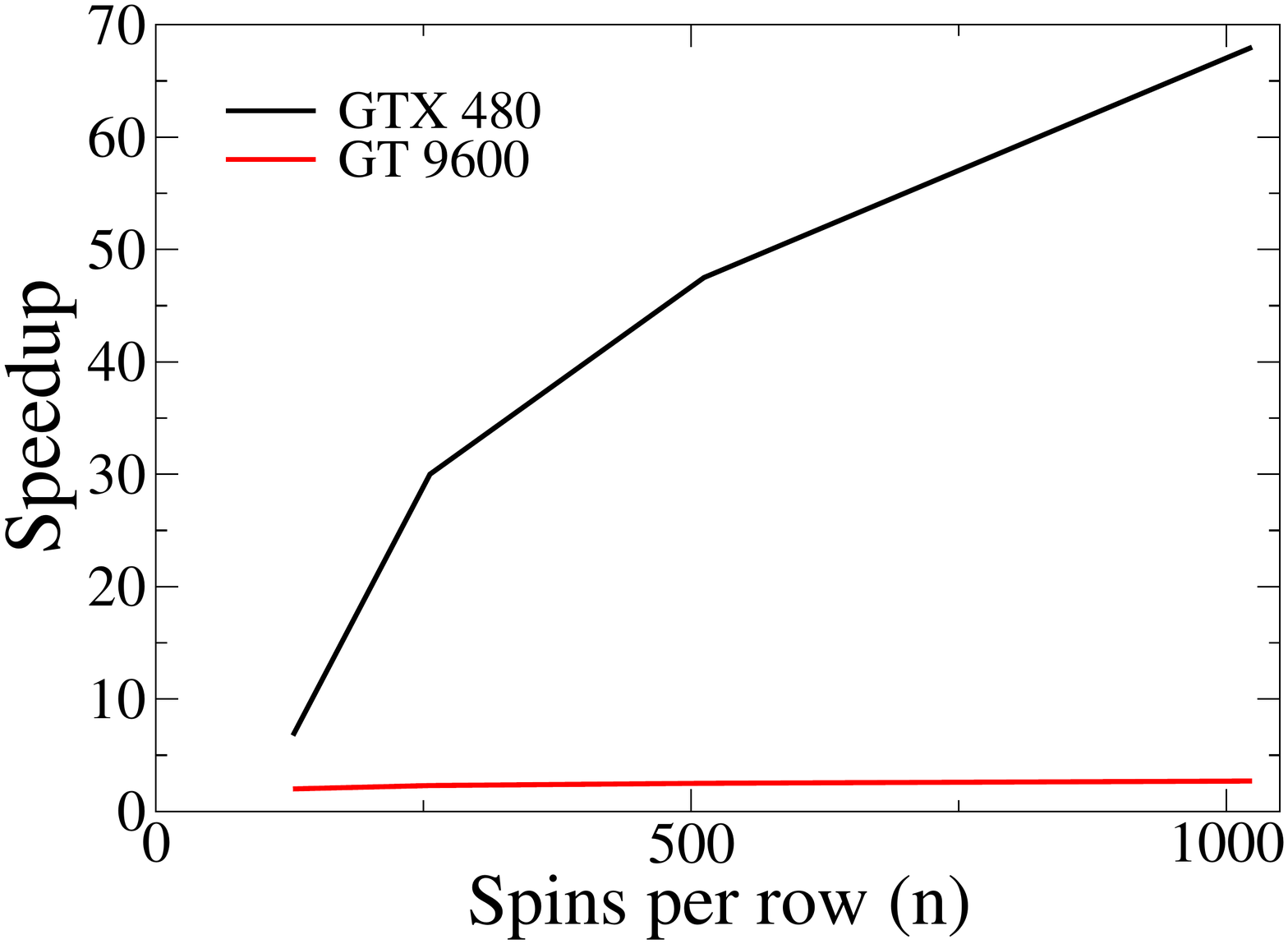}}
\caption{{\small (a) Processing times for the 2D Ising model. Times
are shown as a function of the number of spins $n$ per row which is
related to the system size by $N=n^{2}$. A maximum acceleration factor
of $15$ was achieved (on the GT 9600GPU). We believe a higher value
can be obtained for larger lattices. (b) Speedups of two different GPU
cards versus the CPU serial implementation for the North-East
model. Again times are shown as a function of the number of spins $n$
per row.}}
\label{fig:Running times}
\end{center}
\end{figure}


\section{Conclusions and Summary} \label{sec:conclusions}
We have developed algorithms to simulate lattice spin models on
consumer-grade GPUs. Two prototype models have been considered: The
Ising model describing critical phenomena at equilibrium and the
North-East model describing glassy dynamics. We showed that for
equilibrium properties of lattice models an impressive speedup of
$150\times$ can be achieved. To simulate the dynamics of such models,
a more sophisticated approach was developed in order to preserve the
dynamical rules and outcome. Our algorithm for the dynamic model
reaches a $\approx70\times$ factor in comparison to the serial CPU
implementation for large system sizes. Though the algorithms were
performed on specific models, we feel they can be easily extended to a
larger class of similar systems.\\ Since the gain in computational
power embodied by these results is in some cases two orders of
magnitude, while the required GPU hardware is priced similarly to a
CPU and can often be added to existing systems, the algorithms are
certainly of interest. On the other hand, taking full advantage of it
still requires savvy knowledge of the device and its capabilities and
limitations, and the development of specific numerical algorithms. As
the advantages of this useful technology become clear and knowledge
about its implementation continues to build up and spread, we hope it
will become more accessible to a wide variety of
computationally-minded scientists.

\section{Acknowledgments} \label{sec:acknowledgments}

We would like to thank Prof. Sivan Toledo for discussions. This work
was supported by the Israel Science Foundation (grant no.  283/07). GC
is grateful to the Azrieli Foundation for the award of an Azrieli
Fellowship. ER thanks the Miller Institute for Basic Research in
Science at UC Berkeley for partial financial support via a Visiting
Miller Professorship.

\bibliographystyle{achemso}
\addcontentsline{toc}{section}{\refname}\bibliography{gpu}
\end{document}